\documentclass[preprint]{revtex4}
\usepackage{graphicx}
\usepackage{dcolumn}
\usepackage{bm}

\begin{document}

\preprint{APS/123-QED}

\title{Long-term memory in the Irish market (ISEQ): evidence from wavelet analysis.}

\author{Adel Sharkasi, Heather J. Ruskin and Martin Crane}
\affiliation{%
School of Computing, Dublin City University\\
Email: asharkasi, hruskin and mcrane@computing.dcu.ie
}%

\date{\today}

\begin{abstract}
Researchers have used many different methods to detect the
possibility of long-term dependence (long memory) in stock market
returns, but evidence is in general mixed. In this paper, three
different tests, (namely Rescaled Range (R/S), its modified form,
and the semi-parametric method (GPH)), in addition to a new
approach using the discrete wavelet transform, (DWT), have been
applied to the daily returns of five Irish Stock Exchange (ISEQ)
indices. These methods have also been applied to the volatility
measures (namely absolute and squared returns). The aim is to
investigate the existence of long-term memory properties. The
indices are Overall, Financial, General, Small Cap and ITEQ and
the results of these approaches show that there is no evidence of
long-range dependence in the returns themselves, while there is
strong evidence for such dependence in the squared and absolute
returns. Moreover, the discrete wavelet transform (DWT) provides
additional insight on the series breakdown. In particular, in
comparison to other methods, the benefit of the wavelet transform
is that it provides a way to study the sensitivity of the series
to increases in amplitude of fluctuations as well as changes in
frequency. Finally, based on results for these methods, in
particular, those for DWT of raw (or original), squared and
absolute returns, it can be concluded that there is strong
indication for persistence in the volatilities of the emerging
stock market returns for the Irish data.\\
\textbf{keywords:} {long-term memory, classical and modified R/S
methods, GPH test and the Discrete Wavelets Transform.}
\end{abstract}

\maketitle

\section{\label{sec1}Introduction}

There is no unique definition of long-term memory (\textbf{LTM})
processes (first introduced in \cite{hur51}) which measure long
range dependence. Such a process is generally defined as a series
having a slowly declining correlogram or an infinite spectrum at
zero frequency (see \cite{gra-din-96}).

The existence of long-range dependence (\textbf{LRD}) in a stock
market has been an important topic in recent financial research as
LTM models are able to describe features of data of various
granularities using the same parameters. Several studies find
evidence of long memory in  stock market returns.
(\cite{lee-etal-01}, \cite{sad-sil-01} and \cite{ass-cav-05}),
while others show that there is none or at best there is weak
evidence (\cite{lo91}, \cite{che-lai-95}, \cite{jac96},
\cite{hie-jon-97}, \cite{ber-lyh-98} and \cite{din-etal-93}).

In \cite{cra94}, the author utilized GPH, classical and modified
Rescaled Range (R/S) methods to test the presence of LTM in the
international stock index returns of the G-7 countries and found
no evidence of LTM in these series with the exception of West
Germany. Further, the authors, in \cite{bar-bau-96}, applied the
Spectral Regression Method to test for long-range dependence in 10
U.S stock returns, and returns  of 30 firms included in the Dow
Jones Industrial Index. Their results showed no evidence of long
memory in these returns, as a whole, but some evidence for
persistence in five companies while three other firms exhibited
anti-persistence. In \cite{lob-sav-98}, the authors used a
Lagrange Multiplier procedure and reported that long memory exists
in the squared returns but not in the returns themselves.

More recently, Elekdag \cite{ele01} applied GPH methods to
volatilities of a large data set of emerging markets and found
strong evidence for LTM in these series. This evidence was robust
to various volatilities, specifically the absolute and modified
log-squared returns. Further, Sibbertsen \cite{sib02} found
significant evidence of LTM in the volatilities of several German
stock returns, investigated using the classical and tapered
log-periodogram regression methods. Several articles detail the
fact that emerging capital markets are more likely to have LTM
than the major capital markets (\cite{bek-har-95}, \cite{wri99},
\cite{nat01}, \cite{hen02} and \cite{tol03}).

It seems clear from the literature, therefore, that the
volatilities of stock returns, such as squared and absolute
returns, are more likely to have LTM than the stock returns
themselves. The aim of this paper is twofold; (1) To employ the
discrete wavelet transform (\textbf{DWT}) (for description see,
\cite{lee02}, \cite{sha-etal-05} and \cite{sha-etal-06}) as a new
testing approach to investigate the existence of LTM in the
returns series and volatilities of five Irish indices. (2) To
compare DWT with other methods [namely Rescaled Range (R/S)
(introduced in \cite{hur51}), its modified form (for detail see
\cite{lo91}) and GPH (suggested in \cite{gew-por-83})]. We examine
Irish index returns, (namely Overall, Financial, General, Small
Cap and ITEQ), in order to determine whether long memory behaviour
is exhibited in any or all of these.

The remainder of this paper is organized as follows: In
Section~\ref{DR} the data and results are described and our
conclusion is presented in section~\ref{CON}.

\section{Data and Results}\label{DR}
\subsection{Data Overview}

The data sets considered in this study are the daily closing
values of five Irish Stock Exchange (ISEQ) indices, namely
Overall, Financial, General (from 04/1/1988 to 30/9/2003), Small
Cap (from 4/1/1999 to 30/9/2003) and ITEQ (from 4/1/2000 to
30/9/2003). The daily returns of all these indices are calculated
as follows, [Daily Returns$=\ln (P_t/P_{t-1})$ where $P_t$ and
$P_{t-1}$ are the index price at time $t$ and $t-1$ respectively].

\subsection{Results}
The R/S and Lo's R/S analysis are applied to the index returns and
their volatility measures and the results are reported in
Table~\ref{table1}. There appears to be little evidence of long
memory property in all returns series themselves (from either
method). There is, however, strong evidence of long-range
dependence in the absolute and squared returns of all indices
except in those of the Small Cap index.
\begin{table}[ht]
\caption{\label{table1}Results of the R/S analysis and Lo's
modified R/S test.} {\footnotesize{
\begin{ruledtabular}
\begin{tabular}{cccc}
 ISEQ index&Series &  $V$-test of R/S& $V$-test of Lo's R/S\\
\hline
&Returns&1.7469&1.4776\\
\cline{2-4}
Overall&Absolute&7.2223**& 4.3492**\\
\cline{2-4}
&Squared &5.0001**&3.2941**\\
\hline
&Returns&1.4493&1.3007\\
\cline{2-4}
Financial&Absolute&7.8112**&4.6176**\\
\cline{2-4}
&Squared&5.9294**&3.7573**\\
\end{tabular}
\end{ruledtabular}}}
\end{table}

\clearpage
\newpage

\begin{table}[t]
TABLE~\ref{table1}: Continued. \vspace{1 mm} {\footnotesize{
\begin{ruledtabular}
\begin{tabular}{cccc}
ISEQ index&Series &  $V$-test of R/S& $V$-test of Lo's R/S\\
\hline
&Returns&1.7150&1.4154\\
\cline{2-4}
General&Absolute&7.5615**&4.8269**\\
\cline{2-4}
&Squared&4.2129**&3.0913**\\
\hline
&Returns&1.3499&1.1666\\
\cline{2-4}
Small Cap&Absolute&1.3587&1.1042\\
\cline{2-4}
&Squared&1.4102&1.0757\\
\hline
&Returns&1.7038&1.6499\\
\cline{2-4}
ITEQ&Absolute&2.7325**&1.9878*\\
\cline{2-4}
&Squared&2.1961**&1.6664\\
\end{tabular}
\end{ruledtabular}}}
\footnotetext{\textit{Note: $V$-tests are calculated as
$V_n=W_n/\sqrt{n}$. The acceptance or rejection of the null
hypothesis at $\alpha \%$ level for 5\% or 1\% is determined by
whether or not $V_n$ is contained in the interval [0.809,1.862] or
[0.721,2.098] respectively. Thus * and ** indicate statistical
significance at the 5\% and 1\% respectively.}}
\end{table}

The spectral regression procedure (GPH) is also applied to
estimate $d$ and to test the hypothesis \mbox{($H_0:d=0$ vs.
$H_1:d\not=0$)} for index returns and their volatilities. We
report the GPH test for different values of $\alpha$ =0.45, 0.50,
0.55, 0.60 in order to measure the sensitivity of this test to the
choice of $m$ (where $m=n^\alpha$). 
 A two-sided test is performed, by constructing a $t$-statistic
with the theoretical variance of the spectral regression error
equal to $\pi^2/6$, to test the statistical significant of the $d$
estimates. The results are reported in Table~\ref{table2}. Based
on this analysis, there is no evidence of long-term memory in any
of the returns series and there is strong indication of
persistence in the absolute and the squared returns of all
indices, except that of the Small Cap index. The GPH method shows
that the squared returns of the General index have no long memory
behaviour while both $R/S$ tests show that long-range dependence
is strongly exhibited in this series.

\begin{table}[t]
\caption{\label{table2}GPH estimation of fractional differencing
parameter $d$ for daily returns of Irish Stock Exchange (ISEQ)
indices.} {\footnotesize{
\begin{ruledtabular}
\begin{tabular}{cccccc}
 &&&&$\alpha$&\\
\cline{3-6}
Index$\downarrow$&Series&0.45&0.50&0.55&0.60\\
\hline
&Returns&\textbf{0.0523}&\textbf{0.0519}&\textbf{0.0428}&\textbf{0.1197}\\
& &(0.452)&(0.571)&(0.599)&(1.697)\\
\cline{2-6}
Overall&Absolute&\textbf{0.5060}&\textbf{0.4380}&\textbf{0.4000}&\textbf{0.3650}\\
& &(4.371)**&(4.810)**&(5.600)**&(6.387)**\\
\cline{2-6}
&Squared&\textbf{0.3853}&\textbf{0.3333}&\textbf{0.2966}&\textbf{0.2663}\\
& & (3.256)**&(3.664)**&(4.150)**&(4.665)**\\
\hline
&Returns&\textbf{-0.0019}&\textbf{0.1096}&\textbf{0.1151}&\textbf{0.1289}\\
& &(-0.016)&(1.205)&(1.309)&(1.742)\\
\cline{2-6}
Financial&Absolute&\textbf{0.5313}&\textbf{0.4720}&\textbf{0.3617}&\textbf{0.3384}\\
& &(4.586)**&(5.189)**&(5.060)**&(5.926)**\\
\cline{2-6}
&Squared&\textbf{0.3925}&\textbf{0.3754}&\textbf{0.3250}&\textbf{0.3065}\\
& &(3.388)**&(4.127)**&(4.548)**&(5.368)**\\
\hline
&Returns&\textbf{-0.0280}&\textbf{0.0508}&\textbf{0.0696}&\textbf{0.0972}\\
& &(-0.242)&(0.558)&(0.974)&(1.703)\\
\cline{2-6}
General&Absolute&\textbf{0.3959}&\textbf{0.3615}&\textbf{0.3127}&\textbf{0.3098}\\
& &(3.417)**&(3.974)**&(4.375)**&(5.426)**\\
\cline{2-6}
&Squared&\textbf{0.2188}&\textbf{0.0969}&\textbf{0.0551}&\textbf{0.0877}\\
& &(1.889)&(1.066)&(0.771)&(1.537)\\
\hline
&Returns&\textbf{-0.0874}&\textbf{0.0787}&\textbf{0.0377}&\textbf{0.1369}\\
& & (-0.543)&(0.607)&(0.362)&(1.611)\\
\cline{2-6}
Small Cap&Absolute&\textbf{-0.0189}&\textbf{0.1261}&\textbf{0.1647}&\textbf{0.1586}\\
& &(-0.118)&(0.972)&(1.579)&(1.867)\\
\cline{2-6}
&Squared&\textbf{0.1143}&\textbf{0.0209}&\textbf{0.0278}&\textbf{0.0153}\\
& &(0.710)&(0.161)&(0.267)&(0.180)\\
\end{tabular}
\end{ruledtabular}}}
\end{table}

\clearpage
\newpage
\begin{table}[t]
TABLE~\ref{table2}: Continued. \vspace{1 mm}  {\footnotesize{
\begin{ruledtabular}
\begin{tabular}{cccccc}
 &&&&$\alpha$&\\
\cline{3-6}
Index$\downarrow$&Series&0.45&0.50&0.55&0.60\\
\hline
& Returns&\textbf{0.1011}&\textbf{0.0609}&\textbf{-0.0192}&\textbf{0.0729}\\
& &(0.576)&(0.436)&(-0.170)&(0.786)\\
\cline{2-6}
ITEQ&Absolute&\textbf{0.5370}&\textbf{0.4723}&\textbf{0.4081}&\textbf{0.3169}\\
& &(3.061)**&(3.371)**&(3.621)**&(3.413)**\\
\cline{2-6}
&Squared&\textbf{0.4161}&\textbf{0.3411}&\textbf{0.2989}&\textbf{0.2316}\\
& &(2.372)*&(2.435)*&(2.651)**&(2.495)*\\
\end{tabular}
\end{ruledtabular}}}
\footnotetext{\textit{Note: The $\textbf{d}$  estimates
(\textbf{bold}) corresponding to GPH of $\alpha$. The t-tests are
given in parentheses and their statistical significance are
indicated by * and ** at the 5\% and 1\% significance level
respectively.}}
\end{table}
While these more conventional analyses are useful, serving to
contrast the Irish with other markets' data, we now consider the
relatively novel approach using the discrete wavelet transform
(DWT) to analyze the volatility more directly. The DWT with
symmlet 8 wavelet ($s8$) for 6 levels (scales) is computed for
daily returns series and their volatility measures ( namely
squared and absolute returns) of all Irish indices in order to
investigate the long-term memory property. The DWT provides a more
detailed breakdown of the contribution to the series energy from
the high and low frequencies in the following manner.
Table~\ref{table3} (Panels: A, B and C) display the energy
percentages for wavelet components (crystals) of the returns,
squared and absolute, of Overall, Financial, General, Small Cap
and ITEQ indices respectively. These percentages indicate the
proportion of energy in these series explained by each wavelet
crystal. From Table~\ref{table3} (Panel A), it can be seen that
high-frequency crystals (especially the first and the second) have
much more energy than the lowest frequency one and this means that
movements in the returns are mainly caused by the short-term
fluctuations. This confirms that there is little evidence for long
memory in the returns series.
\begin{table}
\caption{\label{table3}Energy Percentages explained by each
wavelet component.} {\footnotesize{Panel A: The daily returns of
Irish indices. \vspace{1 mm}
\begin{ruledtabular}
\begin{tabular}{cccccccc}
 \textbf{Index $\to$} &\textbf{Overall}&\textbf{Financial}&\textbf{General}&\textbf{Small Cap}&\textbf{ITEQ}\\
\textbf{W.Crystals$\downarrow$} &&&&&\\
\hline
$d_1$&\bf{0.433}&\bf{0.431}&\bf{0.447}&\bf{0.493}&\bf{0.476}\\
\hline
$d_2$&\bf{0.239}&\bf{0.251}&\bf{0.236}&\bf{0.234}&\bf{0.210}\\
\hline
$d_3$ &0.158&0.163&0.138&0.093&0.181\\
\hline
$d_4$ &0.079&0.074&0.083&0.078&0.055 \\
\hline
$d_5$ &0.036&0.033&0.037&0.045&0.036\\
\hline
$d_6$&0.029&0.024&0.029&0.023&0.027 \\
\hline
$s_6$&\bf{0.026}&\bf{0.024}&\bf{0.030}&\bf{0.035}&\bf{0.014}\\
\end{tabular}
\end{ruledtabular}}}
\vspace{4 mm}
{\footnotesize{ Panel B: The squared returns of Irish indices.
\vspace{1 mm}
\begin{ruledtabular}
\begin{tabular}{cccccccc}
\textbf{Index $\to$} &\textbf{Overall}&\textbf{Financial}&\textbf{General}&\textbf{Small Cap}&\textbf{ITEQ}\\
\textbf{W.Crystals$\downarrow$} &&&&&\\
\hline
$d_1$&\bf{0.367}&\bf{0.326}&\bf{0.388}&\bf{0.314}&\bf{0.317}\\
\hline
$d_2$&\bf{0.162}&\bf{0.182}&\bf{0.188}&\bf{0.234}&\bf{0.185}\\
\hline
$d_3$ &0.121&0.115&0.116&0.217&0.125\\
\hline
$d_4$ &0.074&0.059&0.118&0.047&0.069 \\
\hline
$d_5$ &0.046&0.047&0.044&0.042&0.049\\
\hline
$d_6$ &0.026&0.021&0.024&0.031&0.016 \\
\hline
$s_6$ &\bf{0.205}&\bf{0.251}&\bf{0.122}&\bf{0.116}&\bf{0.240}\\
\end{tabular}
\end{ruledtabular}}}
\end{table}
\begin{table}[t]
{\footnotesize{ Panel C: The absolute returns of Irish indices.
\vspace{1 mm}
\begin{ruledtabular}
\begin{tabular}{cccccccc}
 \textbf{Index $\to$} &\textbf{Overall}&\textbf{Financial}&\textbf{General}&\textbf{Small Cap}&\textbf{ITEQ}\\
\textbf{W.Crystals$\downarrow$} &&&&&\\
\hline
$d_1$ &\bf{0.195}&\bf{0.183}&\bf{0.207}&\bf{0.194}&\bf{0.194}\\
\hline
$d_2$&\bf{0.103}&\bf{0.104}&\bf{0.110}&\bf{0.116}&\bf{0.097}\\
\hline
$d_3$ &0.060&0.063&0.062&0.080&0.069\\
\hline
$d_4$ &0.035&0.032&0.045&0.027&0.036 \\
\hline
$d_5$ &0.027&0.027&0.027&0.022&0.024\\
\hline
$d_6$ &0.015&0.013&0.015&0.015&0.008 \\
\hline
$s_6$ &\bf{0.565}&\bf{0.579}&\bf{0.533}&\bf{0.546}&\bf{0.571}\\
\end{tabular}
\end{ruledtabular}}}
\end{table}
Table~\ref{table3} (Panel B) shows that the lowest frequency
component ($s_6$) of the squared returns of each of the Overall,
Financial and ITEQ indices has more energy than the second
high-frequency component ($d_2$) but less energy than the first
crystal ($d_1$). This provides futher detail on previous analysis
and implies that movements in these squared returns are caused by
{\textit{both}} short-term and long-term fluctuations.  Thus there
is  clear evidence of a long memory property in the squared
returns series. The lowest frequency component ($s_6$) of the
squared returns of the General index has lower energy than the
second highest frequency ($d_2$) but higher than that of the third
component, indicative of a {\textit{weak}} long memory effect in
the squared returns of General index. However, the energy of the
lowest frequency crystal of the squared returns of Small Cap index
is even lower than that of the $d_3$ component and this clearly
implies that the movements of this series are mostly caused by
short-term fluctuations with no significant evidence of long-term
memory.

Table~\ref{table3} (Panel C), in contrast, illustrates a situation
where the lowest frequency component ($s_6$) has much more energy
than both the first ($d_1$) and the second ($d_2$) components
together, which is strong evidence of long-range dependence in the
absolute returns series with movements in these series mostly
caused by long-term fluctuations. From the wavelet analysis it is
clear that the frequency patterns are demonstrably different for
the respective series where large energy percentages, associated
with high frequency components, implies short-term memory
dominance and {\textit{vice versa}.
\section{Conclusion}\label{CON}
In this article, the discrete wavelet transform (DWT) and three
other methods, were employed to test for the presence of long
memory in the five Irish Stock Exchange (ISEQ) indices. In
agreement with findings for other indices, (e.g.
\cite{lee-etal-00}, \cite{ele01} and \cite{sib02}), there is no
evidence of long memory for returns series,  while for squared and
absolute returns, such a property does appear to exist. The
exception is the Small Cap index for the Irish data, which shows
no significant evidence of long-term dependence for any returns
series.  The DWT analysis, however, provides additional insight on
the series breakdown. In particular, in comparison to other
methods, the benefit of wavelet transform is that it provides a
way to study the sensitivity of the series to increases in
amplitude of fluctuations  as well as changes in frequency.
Finally,  based on results for these methods, in particular, those
for DWT of returns, squared and absolute returns, it can be
concluded that there is strong indication for persistence in the
volatilities of the emerging stock  market returns for the Irish
data.

\newpage

\end{document}